\documentclass[preprint,amsmath,amssymb,superscriptaddress]{revtex4-2}

\usepackage{graphicx}% Include figure files
\usepackage{dcolumn}% Align table columns on decimal point
\usepackage{bm}% bold math
\usepackage{hyperref}
\begin{document}

\title{Graph Transformer Networks for Accurate Band Structure Prediction: An End-to-End Approach}

\author{Weiyi Gong}
 \affiliation{Department of Physics, Northeastern University, Boston, MA 02115, USA}

\author{Tao Sun}
 \affiliation{Department of Computer Science, Stony Brook University, Stony Brook, NY 11794, USA}

\author{Hexin Bai}
 \affiliation{Department of Computer and Information Sciences, Temple University, Philadelphia, PA 19122, USA}

\author{Jeng-Yuan Tsai}
 \affiliation{Department of Physics, Northeastern University, Boston, MA 02115, USA}

\author{Haibin Ling}
 \email{hling@cs.stonybrook.edu}
 \affiliation{Department of Computer Science, Stony Brook University, Stony Brook, NY 11794, USA}

\author{Qimin Yan}
 \email{q.yan@northeastern.edu}
 \affiliation{Department of Physics, Northeastern University, Boston, MA 02115, USA}

\begin{abstract}

Predicting electronic band structures from crystal structures is crucial for understanding structure-property correlations in complex solid-state materials. First-principles approaches are accurate but computationally intensive. Recently, machine learning (ML) has been extensively applied to this field. Existing ML models predominantly focus on band gap predictions or indirect band structure estimation via solving predicted Hamiltonians. An end-to-end model to predict band structure accurately and efficiently is still lacking. Here, we introduce a graph transformer-based end-to-end approach that directly predicts band structures from crystal structures with high accuracy. Our method leverages the continuity of the k-path and treats continuous bands as a sequence. We propose to apply Fast Fourier Transform to treat continuous bands as sequential signals, which greatly reduces the complexity of modeling and data representations. We demonstrate that our model not only provides accurate band structure predictions but also can derive other properties with high accuracy, such as band gap, band center, and band dispersion. We further show that the model has the capability to generalize from one dataset to another by using transfer learning. Our method has great application potential for the fast prediction of electronic band structures of solid-state materials and automation of material screening, discovery and inverse design.

\end{abstract}

\maketitle

\section*{Introduction}\label{sec: introduction}

The electronic structure of solid-state materials is one of the fundamental properties from which multiple physical properties can be derived. Accurate and efficient electronic structure prediction is crucial for understanding complex material functionalities and enabling the data-driven design of functional materials for technological applications. Electronic structures can be characterized in multiple aspects. The first tier includes single-value quantities such as the effective masses, band gaps for nonmetals, the Fermi energies and work functions for metals. In recent years, the prediction of these single-value properties has been assisted and accelerated by machine learning, more specifically, graph neural network (GNN) models in crystalline systems to characterize their electronic structures \cite{xie2018, chen2019, park2020, choudhary2021, ruff2023}. Despite the incorporation of more complex graph structures and extensive physical information, the capability of these models is limited when exploring more intricate applications where predictions beyond single-value targets are needed. 

More recently, multiple models have been proposed to predict multi-valued properties associated with electronic structures, such as the electronic density of states (DOS). For instance, the study by Yeo \textit{et al.} \cite{yeo2019} treats the DOS as an image-like target that can be learned and subsequently predicted. Mat2Spec \cite{kong2022} employs contrastive learning to encode and align the representations of crystal graph features with the DOS. In a recent work, Xtal2DoS \cite{bai2022} considers DOS as a sequence and employs a graph transformer to encode the crystal graph, followed by a graph-to-sequence (graph2seq) model serving as a decoder to predict the DOS sequence. Although DOS can provide simple and explicit information about the electronic structure, the electronic band structure offers a more detailed and explicit representation of crystal properties. Therefore, a machine learning model capable of predicting electronic band structures has become increasingly desirable in recent years.

As an explicit and comprehensive representation of electronic structures, electronic band structures are obtained from first principles calculations which are expensive for complex material systems. For this reason, the prediction of band structures from crystal structures is considered one of the most important tasks in the application of ML to solid-state physics and the data-driven design of functional materials. Recently, an indirect approach has been proposed to train a machine learning model to predict the tight-binding Hamiltonian matrix elements, followed by solving the eigenvalues of the predicted Hamiltonian to obtain the electronic band structure \cite{li2022,zhong2023}. The approach has been applied to multiple complex material systems, including twisted bilayer graphene and twisted bilayer bismuthene \cite{li2022} and SiGe alloy systems \cite{su2023}. In a more recent model, DeepH-E3 \cite{gong2023} integrates GNN with equivariant neural networks (ENN) \cite{cohen2016, thomas2018, geiger2022}, which incorporates specific gauge symmetries in three dimensional space, enabling more data-efficient and accurate model training. However, the investigated material systems are restricted to some specific types of material systems. Hence, a universal model that can be applied to large-scale material datasets is yet to be proposed. Moreover, the need to obtain the eigenvalues of the predicted Hamiltonian with eigenvalue solver limits the scalability and efficiency of these models, thus posing challenges to their practical application in high-throughput material screening. 

To address these challenges, we introduce the first end-to-end model \textit{Bandformer} to predict band structures directly from crystal structures. The model is designed based on the architecture of transformer \cite{vaswani2017}, Graphormer \cite{ying2021} and Xtal2DoS \cite{bai2022}. We treat the complex relationship between atomic interactions, high symmetry k-paths in reciprocal space, and band energies as a "language translation" task. The crystal structure is encoded into a hidden representation and "translated" into the band structure using a graph2seq module. Fast Fourier Transform (FFT) is applied to treat continuous bands as sequential signals, which greatly reduces the complexity of modeling and data representations. The model is trained and tested on a large and diverse dataset consisting of 27,772 band structures from the Materials Project database \cite{jain2013}. We focus on predicting multiple energy bands around Fermi level and achieve a mean absolute error (MAE) of 0.304 eV for band energy prediction. We further demonstrate our model's outstanding performance in predicting properties derived from the band structures, such as band gap prediction for nonmetals with a MAE of 0.251 eV. Our end-to-end model holds great potential in accelerating the data-driven discovery and inverse design of functional materials with specific features associated with electronic band structures. 

\section{Results \label{sec: results}}

\subsection{Definition of band structure learning problem} 

Solving the band structure of crystalline systems is a central problem in solid-state physics, since it is essential for the understanding of structure-property correlations in solids. Traditionally, \textit{ab initio} methods such as density functional theory (DFT) address this by solving the Kohn–Sham eigenvalue equations. For a typical \textit{ab initio} calculation of band structure, the crystal structure, atomic potentials, and k-point set are needed as input. Self-consistent and non-self-consistent calculations are performed to determine the eigenvalues at each predefined k-point, from which the band structure for a given crystalline system can be obtained. Fundamentally, DFT-based band structure calculation represents a highly complicated mapping between a crystal structure and its electronic bands, involving multiple layers of theory and numerical algorithms. According to the universal approximation theorem \cite{cybenko1989, hornik1989, chen1995}, carefully designed ML models, when informed by domain knowledge and trained on large datasets, can statistically approximate such complex mappings with sufficiently low error. Hence, the central question raised and studied in this work is whether a high-quality machine-learned mapping from crystal structure to band structure can be constructed.

The band structure is defined as a piecewise continuous function along high-symmetry paths in k-space. A critical challenge for ML-based band structure prediction is that both the number of bands ($N_b$) and k-points ($N_k$) vary across different materials depending on their crystal symmetry and electronic configuration. This variability prevents direct application of standard ML models that require fixed-size inputs. We address these challenges through two key strategies. Firstly, we focus on a fixed number of bands nearest to the Fermi level, as these are most relevant for determining material properties. Second, we employ a resampling scheme that converts variable-length k-paths into fixed-dimension representations suitable for ML processing. Our approach utilizes continuous Eulerian k-paths to ensure smooth and non-redundant sampling of the Brillouin zone, combined with interpolation techniques to achieve consistent input dimensions across all materials.
This framework enables the construction of a machine-learned mapping from crystal structure to band structure while maintaining computational efficiency and physical accuracy. The specific implementation details of our band selection and k-point resampling algorithms are described in the Methods~\ref{sec: methods} section.

\subsection{The architecture of Bandformer}

\begin{figure*}
\centering
\includegraphics[width=\textwidth]{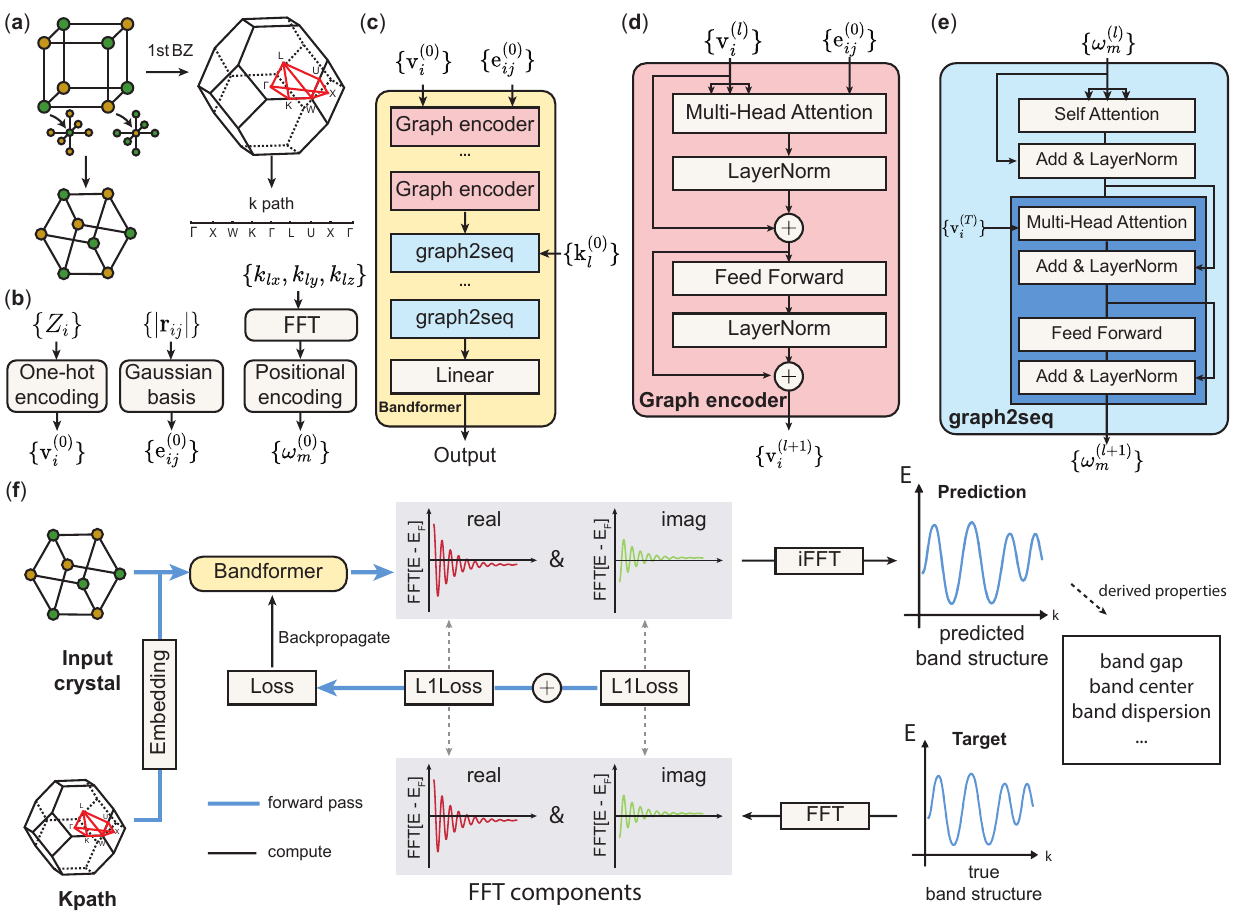}
\caption{
\label{fig:fig1} \textbf{The workflow and architecture of Bandformer.} 
\textbf{(a)} For each input crystal, a crystal graph is constructed from the local environment of atoms in the primitive cell, which is used as the input for the encoder. The high symmetry k-path in the first Brillouin zone is extracted and resampled and used as the input for the decoder graph2seq. 
\textbf{(b)} Atoms are given element features attached to graph nodes, and interatomic distances are given Gaussian expansions of the distance as features.  K-points are given spatial positional encoding. 
\textbf{(c)} The architecture of Bandformer. 
\textbf{(d)} The architecture of graph encoder. The input atomic features and the edge features are used to generate a bias term when calculating the attention coefficients in the Biased Multi-head Attention module, which is connected to LayerNorm layer in a ResNet structure \cite{he2016}. 
\textbf{(e)} For each k-point along high symmetry k-path, positional encoding of k-point coordinates is used as the initial feature, which is passed to a graph2seq decoder. Within the decoder, k-points learn positional information through Self Attention module. The output is passed to the graph2seq attention module, together with the updated graph encoder output.
\textbf{(f)} The workflow of \textit{Bandformer}. The model predicts real and imaginary components of the FFT-transformed band structure, 
}
\end{figure*}

Fig.~\ref{fig:fig1} shows an overview of Bandformer model workflow and architecture of sub-modules. The essence of our model is to treat the band energy prediction task as a language translation task, where the input crystal graph as the input “sentence” is “translated” into the sequence of band energy values as the output “sentence”.  As shown in Fig.~\ref{fig:fig1} (a), for an input crystal, a typical crystal graph is constructed from local environment of atoms in primitive cell. Each node represents an atom, and each edge represents the interatomic distance. Each node is then assigned features using one-hot encoding based on its atomic number. Each edge is assigned features defined by Gaussian expansion, which has been commonly used in GNN applied in crystal structures such as CGCNN \cite{xie2018} and SchNet \cite{schutt2018}. The continuous k-path following the LM scheme is used as the input for decoder, which will run self attention to learn positional information between k-points and \textit{graph2seq} attention to involve updated crystal graph information. Finally, the predicted band structure is obtained by an output multi-layer perceptron (MLP) layer.

As shown in Fig.~\ref{fig:fig1} (c), the crystal graph is passed to a graph transformer encoder to extract hidden representations of nodes by exchanging and updating the graph representations. Edge features are treated as an addition bias term to calculate attention coefficients, similar to the spatial encoding used in Graphormer \cite{ying2021}. The output of biased multi-head attention module and feed forward module are passed to a skip connection structure known as ResNet \cite{he2016}, which is extensively used in deep neural network models and is one of the most fundamental model architectures. LayerNorm \cite{ba2016} is also used to ensure stable results during training.

The architecture of the decoder is shown in Fig.~\ref{fig:fig1} (e). Inspired by the positional encoding of real space coordinates, which was used for graph transformer for crystals \cite{cui2023}, we apply positional encoding of k-points as the initial feature followed by a real Fast Fourier Transform (rFFT) module for the target sequence in the decoder. Details can be found in Methods~\ref{sec: methods} section. The data is then processed by the self-attention module, which helps to identify the relationships among k-points in the sequence. This information is then passed to a \textit{graph2seq} attention module, combined with the graph nodes from the last layer of graph transformer to calculate \textit{graph2seq} attention coefficients. Finally, the decoder output is passed to a MLP layer to generate the predicted band structure along the specified k-path.

The band structure along k-path is treated as a $N_b$ by $N_k$ dimensional data with real values, for which FFT can be used to extract importance features, especially the oscillations of bands. Generally, for a complex-valued one dimensional input, FFT computes:

\begin{equation}
    X[k] = \sum_{n=0}^{N-1} x[n] \cdot e^{-i \frac{2\pi k n}{N}}
\end{equation}
where $n \in [0, N-1]$ is the time \/ real domain index, $k \in [0, N-1]$ is the frequency \/ reciprocal domain index.

The Fourier transform is particularly well-suited for band structures because electronic bands have oscillatory patterns that directly reflect the underlying crystal symmetry and electronic interactions modulated by crystal periodicity. By transforming to "frequency" space, we can efficiently capture both long-range trends and fine-scale variations in the band dispersion, which are difficult to model directly in k-space. Note that the "frequency" is representing only the conjugate quantity of k-point indices along the continuous k-path. In this work, since we use continuous k-point selection scheme, oscillating behavior in most bands are constrained and the general dispersion of most bands can be largely captured by small "frequency" components. 

In this work, since the band energies are real values, we use real Fast Fourier Transform (rFFT) for transforming the band energies. By removing the duplicate symmetric components, the frequency domain indices are constrained to the range $k \in [\lfloor N/2 \rfloor]$.

% As an ablation study, we also divide the original task of predicting band energies into two sub-tasks and trained a model to predict the mean values (band centers) and the difference between band energies and band mean values (band dispersions) separately. Specifically, the total loss is defined as the weighted sum of the MAE loss for the predicted band centers and the MAE loss for the predicted band dispersions, formulated as $\mathcal{L} = \mathcal{L}_1 + \lambda \mathcal{L}_2$, where $\mathcal{L}_1$ and $\mathcal{L}_2$ are the MAE loss for predicted band centers and band dispersions correspondingly, and the weighting factor $\lambda$ is found by hyper-parameter searching algorithm.

\subsection{Data preparation} 

We train and test our model based on one of the largest publicly available datasets of band structures, the Materials Project dataset. The band structures in this database are calculated using Perdew-Burke-Ernzerhof (PBE) functional \cite{perdew1996} or PBE+U method \cite{cococcioni2005}. We select a subset of this dataset which includes the band structures for 27,772 unique non-magnetic solid-state materials. The dataset contains a wide variety of crystal structures, with unit cells varies from 2 to 200 atoms, number of bands varies from 2 to 300, and the number of k-points for the band structures varies from 50 to 2000 points. This diversity provides a robust foundation for model generalizability, enabling the trained model to perform effectively on previously unseen data.

To address the issue of unfixed $N_b$, we select a fixed number ($N$) of bands closest to the Fermi level, which are typically the most relevant for determining material properties. For any given crystal, we identify the highest occupied band below the Fermi level and select $N/2$ bands below and $N/2$ bands above this level. To have a consistent and robust definition of band energy, Fermi level is defined as the calculated one for any metal, while shifted to the middle of the band gap for any insulator. This approach ensures that the most critical electronic states are consistently studied across different materials, while maintaining a fixed $N_b$ value.

To address the issue of unfixed $N_k$, the number of k-points sampled along the high-symmetry path, one may consider padding all to the same length. This is impractical for band structure predictions, where the number of k-points in our dataset can reach up to 1500. Such padding would introduce the computational cost significantly, since the attention mechanism in transformer-based models scales as $\mathcal{O}(S^2)$ \cite{vaswani2017}. Hence a smooth resampling of the k-paths is needed. The standard k-path selection in band structure calculation is based on the Setyawan-Curtarolo (SC) scheme \cite{setyawan2010}. However, this scheme introduces discontinuity in k-paths to avoid repeating, which will make use of discontinuous data and cause unstable predictions. To generate continuous and moreover, Eulerian k-paths, we use graph theory to add connections between k-points of odd-degree, as is proposed in the Latimer-Munro (LM) scheme \cite{munro2020}. By using this algorithm, the k-paths for band structures are continuous, which ensures the smoothness of bands along these paths and produces smooth predictions.  Additionally, to ensure a fixed number of k-points, we first apply a smoothing Gaussian filter function to interpolate the original data points and then resample the band structure using a predetermined number of k-points. The number of resampled k-points is set to 128 for all crystal structures in this work. This process requires a continuous k-path, which is why we use the LM scheme instead of the SC scheme. Although the LM scheme will have redundant k-points, it removes discontinuities in the high-symmetry path, making it suitable for interpolation and resampling. To focus on the bands of greatest interest, we first identify the highest nonempty band and select $N/2$ bands above and $N/2$ bands below it, including itself. $N$ is set to 6 throughout this work and can easily be increased for predicting more bands. The Fermi level remains unchanged for metals, while for non-metals, it is shifted to the middle of the band gap. This ensures more stable training and evaluation of the model. Additional details are provided in the Supplementary Information. 

The decoder in our model is based on the transformer architecture. Since transformer uses LayerNorm, which applies standard normalization on each band individually, this approach predicts bands with similar value ranges. This makes transformer more suitable for predicting distribution-like targets, such as phonon DOS and electron DOS. However predicting band structure is more challenging since different bands have different band centers. To address this issue, we proposed and studied two methods in this work: (1) apply FFT to truncated and smoothed band structures and predict the real and imaginary components to reconstruct the band structures (2) decompose the target into two sub-tasks to predict the band centers and the band dispersions separately. Specifically, the model predicts the mean value (band center) and the deviation from the mean (band dispersion) for each band. The final band structure is then reconstructed by combining these components. 

\begin{figure*}
\includegraphics[width=\textwidth]{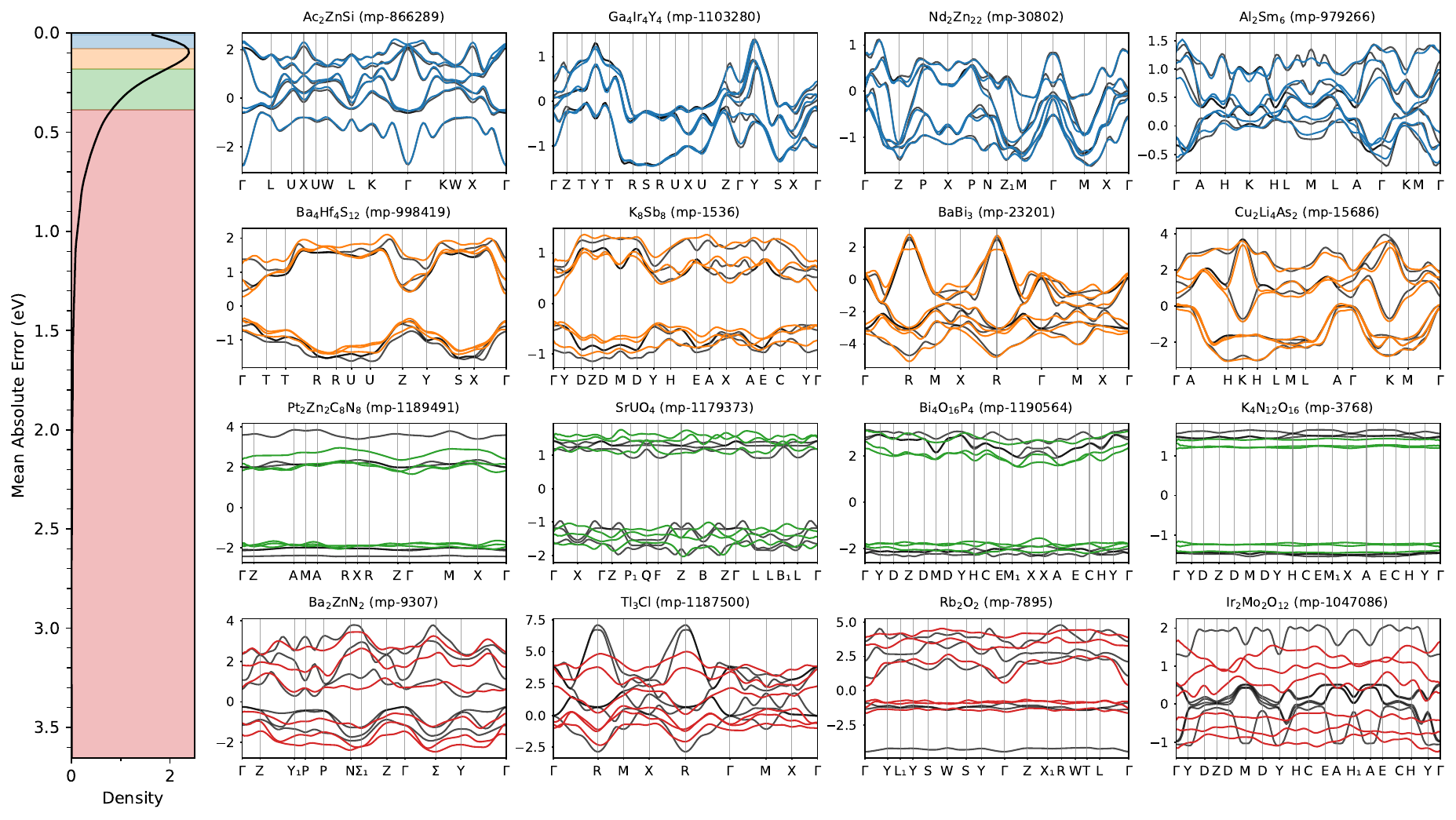}
\caption{\label{fig:fig2}
\textbf{Performance of Bandformer on the test set of Materials Project band structure dataset.} 
Some predicted band structures (colored) vs target band structures (gray) in four quartile of prediction MAE: 0 - 25\% (blue), 25 - 50\% (orange), 50 - 75\% (green), 75 - 100 \% (red). The first quartile has MAE between 0.011 and 0.077 eV, the second quartile has MAE between 0.077 and 0.183 eV, the third quartile has MAE between 0.183 and 0.399 eV and the last quartile has MAE between 0.399 and 3.483 eV.
}
\end{figure*}

\begin{table}
\caption{\label{tab:tab1}
\textbf{Model performance on the test set.} 
Performance is evaluated by MAE for band structure prediction, with Bandformer model, Bandformer with FFT frequency truncation (Bandformer-T) and the transfer-learned model from small cell to large cells.
}

\begin{tabular}{c|c}
\hline\hline
Model  & MAE (eV)\\
\hline
Bandformer     & 0.304 \\
Bandformer-T & 0.306 \\
Bandformer Transfer Learning & 0.419 \\
\hline\hline
\end{tabular}
\end{table}

As shown in Fig.~\ref{fig:fig2}, we visualize the predicted band structures across four quartiles, from $Q_1$to $Q_4$. The total test set is divided into four groups of equal size, sorted by their MAE loss. $Q_1$ represents the group with the lowest error ranging from 0 to 25\% among the whole test set, while $Q_4$ corresponds to the group with the highest error, ranging from 75 to 100\% among the whole test set. Three representative examples from each group is displayed in the figure to illustrate the model's performance across different error ranges. The results demonstrate that our model can predict electronic band structures with high accuracy using an end-to-end approach, without relying on intermediate quantities. The band structure is generated directly from the model outputs, a task that, to the best of our knowledge, has not been previously achieved.

In addition to the direct band structure prediction, other quantities related to bands can also be derived from the predicted band structures. The conduction band minimum (CBM) and the valence band maximum (VBM) can directly be read out from the predicted bands, and band gap can be calculated by the difference between the CBM and the VBM. Also, band center can be calculated from the mean value of each band. And band dispersion can be calculated from the difference between the minimum and the maximum of each band. Moreover, from the predicted band structure, we can directly identify whether a material is metallic or non-metallic. For all materials in the test set, we achieve an MAE of 0.205 eV for band gap predictions, which demonstrates the model’s capability in this side task. The performance of band gap prediction is shown in Fig.~\ref{fig:fig4} (d). Additionally, we achieve band centers and band dispersions are  as shown in  Fig.~\ref{fig:fig4} (e), (f), The performance is summarized in Table.~\ref{tab:tab2}

\begin{table}
\caption{\label{tab:tab2}
\textbf{Predicted properties from predicted band structure.} 
The predicted band gap, band center and band dispersion are calculated directly from predicted band structure. 
}

\begin{tabular}{c|c}
\hline\hline
Property  & MAE (eV)\\
\hline
Band gap     & 0.205 \\
Band center & 0.264 \\
Band dispersion & 0.262 \\
\hline\hline
\end{tabular}
\end{table}

\begin{figure*}
\centering
\includegraphics[width=\textwidth]{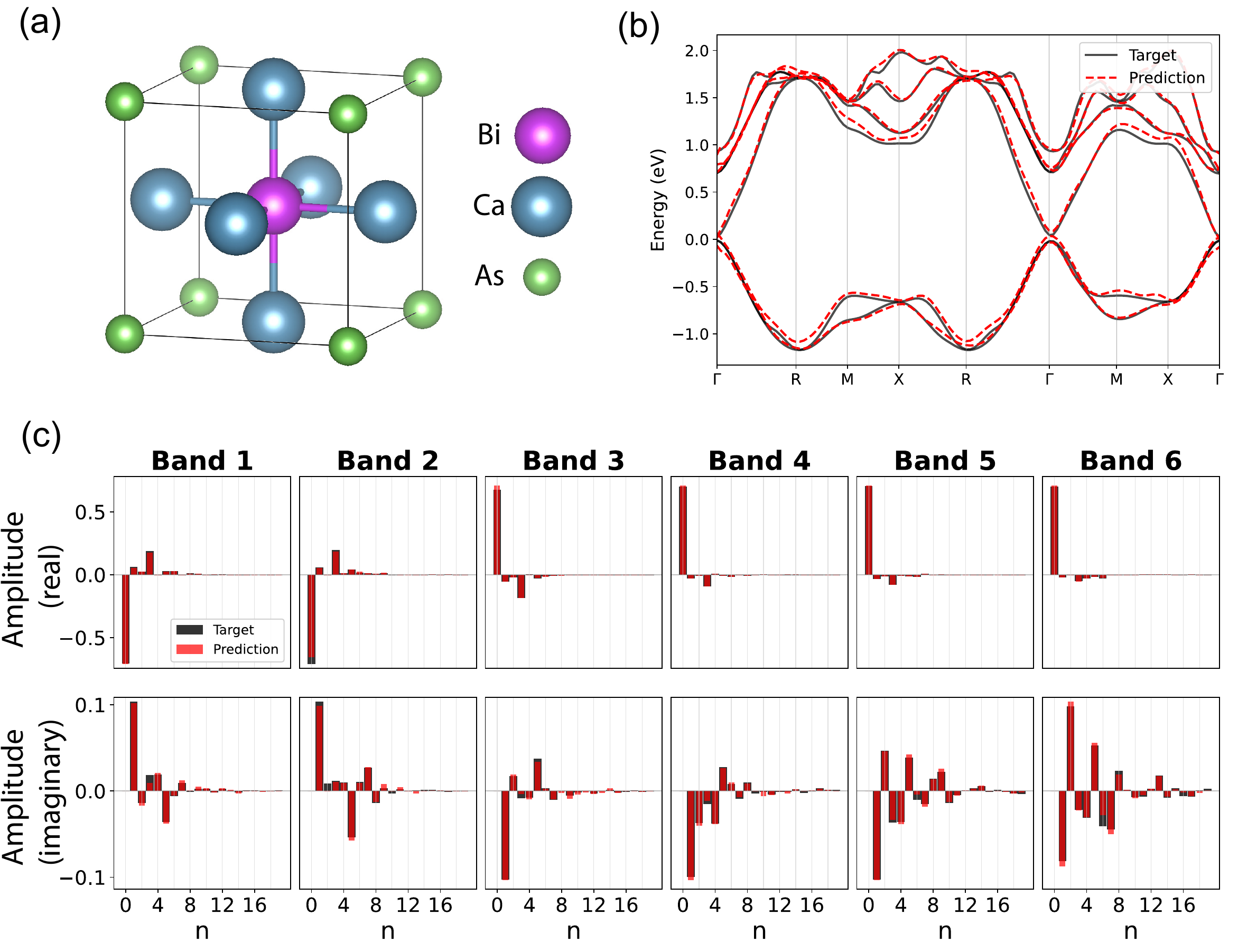}
\caption{
\textbf{An example material from test set.} 
The material has formula Ca$_3$BiAs and Materials Project id mp-1013703. 
\textbf{(a)} The crystal structure of the material.
\textbf{(b)} The predicted vs target band structure of the material.
\textbf{(c)} The predicted vs target FFT components of the material.
}
\label{fig:fig3}
\end{figure*}

\begin{figure*}
\centering
\includegraphics[width=\textwidth]{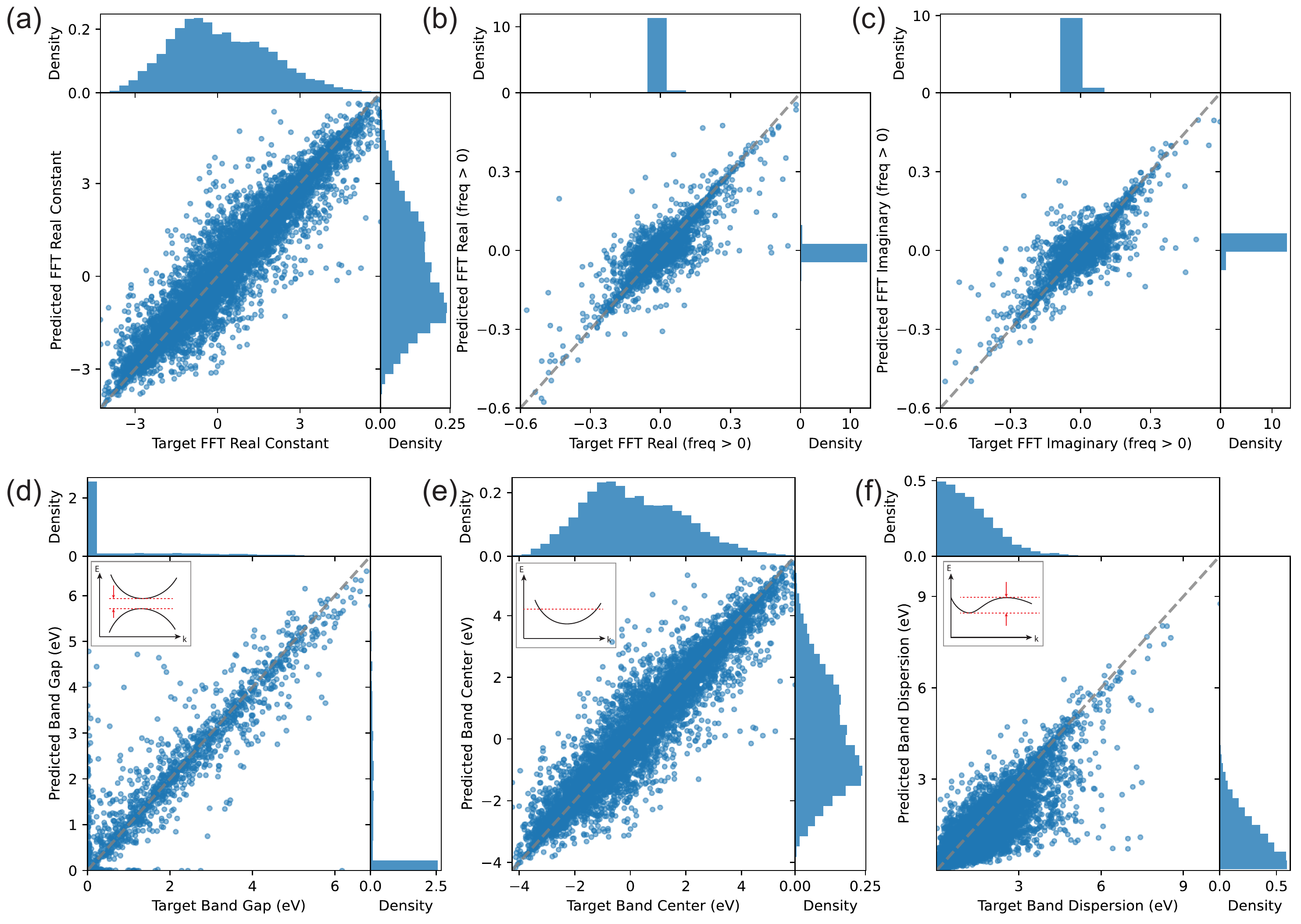}
\caption{
\textbf{The prediction vs target plot for multiple properties:} 
\textbf{(a)} constant term of FFT component;
\textbf{(b)} FFT real components with $\omega > 0$;
\textbf{(c)} FFT imaginary components with $\omega > 0$;
\textbf{(d)} band gaps;
\textbf{(e)} band centers;
\textbf{(f)} band dispersions.
}
\label{fig:fig4} 
\end{figure*}

In Fig.~\ref{fig:fig3}, we show an example material in the test set and how Bandformer can predict its band structure. As shown in Fig.~\ref{fig:fig3} (a), the material is Ca$_3$BiAs with Materials Project id mp-1013703. Our model predicts the FFT components of band structures for the 6 bands around the Fermi level. With a continuous k-path, the band structures along it will be smooth, so that bands will be smooth and have rapidly decaying FFT component magnitudes, which is clearly shown in Fig.~\ref{fig:fig3} (c). The predicted band structure can then be reconstructed from predicted FFT components. 

\section{Discussion \label{sec: discussion}}

In this section, we will explore additional results, limitations of our model, and potential future directions for its development. Firstly, it is important to note that our model has limitations, including its inability to predict an unfixed and larger number of bands without prior user selection. To predict a variable number of bands, one approach could be to not fix $N_b$ and identify the maximum $N_b$ value, supplementing with empty virtual bands for materials with fewer actual bands. A similar method could be applied to the number of k-points. For the Materials Project dataset we used, the maximum values for $N_b$ and $N_k$ could be as high as 200 and 1500, respectively. Predicting a 200 by 1500 matrix would require substantial computational resources that exceed our current capabilities, but it is not unfeasible. 

In this work, we limit training to six bands; however, this number can be easily increased by adjusting the model's hyperparameters and retraining. The current framework allows for straightforward extension to predict more bands without significant architectural changes. The training cost scales linearly with the number of bands, as the decoder transformer maintains the same output shape, with only the number of neurons in the output MLP layer adjusted to match the required band count. This approach ensures linear time complexity with respect to the number of bands, under the assumption that the distributions of different bands are similar or can be learned from the same underlying data distribution. For a more precise modeling of individual band distributions, a potential improvement would involve using $N_b$ separate graph2seq decoders, each dedicated to learning the unique distribution of a specific band. However, this approach would scale with a complexity of $\mathcal{O}(N_b^2)$, significantly increasing computational costs. Given the strong performance of our current model, we opt to retain the existing architecture, which balances accuracy and efficiency.

Another promising direction for future research involves the extraction of descriptors or targets for band structure-related properties that have more physical meaning, which could enable faster training and enhance model accuracy. In our current work, we applied FFT to convert raw band structure data from reciprocal space to an auxiliary "frequency" space, treating the reciprocal space as "time" space. This transformation produces targets with much smaller dimensions, which helps to reduce oscillations and overfitting in our model predictions. 
% However, the exact physical meaning of this auxiliary space remains unknown, though it may be related to the localization of wave function in real space. Developing a low-dimensional descriptor for crystal band structures that can both decode from and encode into the band structure with minimal information loss would be extremely beneficial. Such a descriptor would simplify the representation of complex band structures, making it easier for models to learn and predict these structures accurately. This approach could significantly improve the performance of models designed for predicting band structures.
Looking ahead, developing a large-scale model based on transformer architecture to handle extensive band structure datasets is a promising research direction. Once such a model is trained, it can be directly deployed to make predictions and reduce the computational cost for DFT calculations or Hamiltonian solving for predicting band structures of unknown materials. 
% use for inverse design in the future

In conclusion, we have introduced Bandformer, the first end-to-end model that directly predicts electronic band structures from crystal structures. Our approach treats band structure prediction as a sequence translation task, using graph transformers to encode crystal structures and a graph2seq decoder to generate band energies along high-symmetry k-paths. By applying Fast Fourier Transform to band structures, we achieve efficient representation and stable training while maintaining physical accuracy. The model demonstrates strong performance on a diverse dataset from the Materials Project, accurately predicting both band energies and derived properties such as band gaps, band centers, and band dispersions. This work establishes that machine learning can effectively approximate the complex quantum mechanical mapping between crystal and electronic structures, enabling rapid electronic structure prediction for high-throughput materials screening and accelerating the discovery of functional materials with desired electronic properties.

\section{Methods \label{sec: methods}}

\subsection{Graph transformer encoder}
For a given input crystal structure, we construct a crystal graph from the atoms in the primitive cell and their neighboring environment, selecting atoms within a predetermined searching radius. We consider 12 nearest neighbors within 8 {\AA} searching radius of each atom to create the crystal graph for each input crystal structure. In the encoding phase, the crystal structure is initially represented as a graph $G$, where nodes represent atoms, and edges represent interatomic distances. Each vertex $i$ is characterized by a feature vector $\bm{x}_i^0$ based on properties of the atom such as group number, period number and atomic radius. The performance of model using this type of features is tested and compared to that using just one-hot encoding of atomic number as input features. Each edge $ij$ is represented by a feature vector $\bm{e}_{ij}$, which is a Gaussian function expansion of interatomic distances, which has been used frequently in previous works. We use a graph transformer architecture within MPNN framework \cite{gilmer2017}. At each updating step $l$, edge feature $\bm{e}_{ij}$ and atom feature $\bm{x}_i^l$ are passed to linear layers without bias to generate edge embedding, query, key and value vectors: $\bm{m}_{ij}^l = W_e^l \bm{e}_{ij}$, $\bm{q}_i^l = W_q^l \bm{x}_i^l$, $\bm{k}_i^l = W_k^l \bm{x}_i^l$, $\bm{v}_i^l = W_v^l \bm{x}_i^l$, where $\bm{e}_{ij}$, $\bm{x}_i \in \mathbb{R}^d$ are $d$ dimensional feature vectors and $W \in \mathbb{R}^{d \times d}$ are $d\times d$ dimensional projection matrices.

Neighboring atoms $\mathcal{N}_r(i)$ within a certain radius $r$ of the center atom $i$ are gathered to calculate the attention coefficients by using \textit{softmax} function defined as 

\begin{equation}
    \alpha_{ij}^l 
    % = \mathrm{softmax} \left( \langle \bm{q}_i^l, \bm{k}_j^l; \bm{m}_{ij}^l \rangle \right) 
    = \frac
    {
        \exp{ \left( \langle \bm{q}_i^l, \bm{k}_j^l; \bm{m}_{ij}^l  \rangle \right) }
    }
    {
    \sum_{p \in \mathcal{N}_r(i) } \exp{\left( \langle \bm{q}_i^l, \bm{k}_p^l; \bm{m}_{ip}^l  \rangle \right)}
    },
\end{equation}
% number of heads
where $\langle \bm{q}, \bm{k}; \bm{m}  \rangle = \frac{1}{\sqrt{d}} \bm{q}^T  \bm{k} + \bm{m}$ is a modified multi-head scaled dot product with edge feature projection $\bm{m}$ included. The node features are updated by a skip connection first introduced in ResNet \cite{he2016}: $\bm{x}_i^{l+1} = \bm{x}_i^{l} + \text{LayerNorm} \left( \sum_{j \in \mathcal{N}_r(i) } \alpha_{ij}^l \bm{x}_j^l \right )$. After $L$ iterations, a crystal feature is obtained by the global pooling of  graph nodes $x^L = \sum_{i \in G} x_i^L$. This feature serves as the input for the graph2seq decoder. 

\subsection{Reciprocal space positional encoding}

In decoding phase, we utilize a graph2seq model similar to Xtal2DoS, which has been used to predict electronic DOS. \cite{bai2022} The output from the encoder is utilized to generate key and value vectors. Each k-point along the k-path has coordinates $(k_{ix}, k_{iy}, k_{iz})$. We use the features for the k-point coordinates by passing k-point coordinates to the positional encoding defined as:

\begin{equation}
    f_n(k_i) =\exp \left ( j2\pi \omega_n k_i \right ) \\
\end{equation}
where
\begin{equation}
    \omega_n = \frac{1}{10^{n/(\text{dim}-1)}}, \quad \text{dim} = d_{\text{model}} / 6
\end{equation}
The feature for a k-point is

\begin{equation}
\begin{aligned}
    [ 
        & \cos{(2\pi\omega_0k_x)}, ..., \cos{(2\pi\omega_{\text{dim}-1}k_x)}, \sin{(2\pi\omega_0k_x)}, ..., \sin{(2\pi\omega_{\text{dim}-1}k_x)}, ..., \\
        & \cos{(2\pi\omega_0k_y)}, ..., \cos{(2\pi\omega_{\text{dim}-1}k_y)}, \sin{(2\pi\omega_0k_y)}, ..., \sin{(2\pi\omega_{\text{dim}-1}k_y)}, ..., \\
        & \cos{(2\pi\omega_0k_z)}, ..., \cos{(2\pi\omega_{\text{dim}-1}k_z)}, \sin{(2\pi\omega_0k_z)}, ..., \sin{(2\pi\omega_{\text{dim}-1}k_z)}, ...
    ].
\end{aligned}
\end{equation}

The similar form has been applied to real space coordinates in previous work and proven to be helpful in creating features for coordinates like input \cite{cui2023}. This initial feature is used for generating the query vector for each k-point along the k-paths. $x_i^L$ is used for creating key and value vectors. The initial k-point features, alongside the output from the graph transformer decoder, are iteratively fed into the decoder for multiple time steps. Finally, the final output from the last layer is processed through a multi-layer perceptron (MLP). 

\subsection{Real Fast Fourier Transform Representation}

In this work, we apply a real Fast Fourier Transform (rFFT) to the positional-encoded sequence to compress the information for denser representation of band structures.  After constructing the high-symmetry k-path and resampling to a fixed length of $N_k = 128$, each k-point coordinate $(k_x, k_y, k_z)$ is first mapped into a $d_{\text{model}} = 256$ dimensional vector using the reciprocal space positional encoding described in the previous subsection. This results in an encoded representation of shape $(N_k, d_{\text{model}}) = (128, 256)$.

We then apply the discrete rFFT implemented in PyTorch (\texttt{torch.fft.rfft}) along the $N_k$ dimension. The rFFT converts the input sequence from one dimensional subspace of reciprocal space into frequency space, producing an output of shape $(N_k/2+1, d_{\text{model}}) = (65, 256)$. Each column in this transformed representation encodes the amplitudes and phases of the oscillatory modes for a given embedding dimension.

Using rFFT provides several advantages. Firstly, the band structure values under LM scheme along continuous k-paths are smooth, leading to fast-decaying Fourier magnitudes. This property allows the model to concentrate on low-frequency modes that capture the essential trends of band dispersion. Second, the frequency-domain representation constrains the range of values, which stabilizes optimization and improves convergence during training. Third, rFFT effectively compresses the sequence representation from 128 to 65 points without significant information loss, reducing both computational cost and overfitting risk.

During model training, Bandformer predicts the real and imaginary components of the rFFT-transformed band structures. The predicted frequency-domain representation is then converted back to real space using the inverse rFFT, ensuring that the reconstructed band structures are directly comparable to those obtained from first-principles calculations.

\section{Data availability}
The processed band structure data for model training used in this study will be made available after publication of this work.

\section{Code availability}
The code used in this study is available at \url{https://github.com/qmatyanlab/Bandformer}.

\section{Competing interests}
The authors declare no competing interests.

\begin{acknowledgments}

This work is supported by the U.S. Department of Energy, Office of Science, Basic Energy Sciences, under Award No. DE-SC0023664. This research used resources of the National Energy Research Scientific Computing Center (NERSC), a U.S. Department of Energy Office of Science User Facility located at Lawrence Berkeley National Laboratory, operated under Contract No. DE-AC02-05CH11231 using NERSC award BES-ERCAP0029544.

\end{acknowledgments}

\bibliography{ref}

\end{document}